*"...les enseignants de tous rangs sont inclinés à trouver dans un misonéisme défensif une manière d'échapper au déclassement, et il n'est pas rare qu'ils abusent de la situation de monopole qu'assure l'enseignement pour prendre de fausses distances à l'égard de savoirs qu'ils auraient en tout cas peine à transmettre."*

P. Bourdieu, *Homo Academicus* (1984).

# UNDERSTANDING THE GENESIS OF MASS:
## A mystery pointing towards the mirror world

*George Triantaphyllou*
*Physics Department, National Technical University of Athens*
*Zografou Campus, GR 157 80 Athens, Greece*

September 5, 2002

**Abstract**

We give a pedagogical and concise presentation of dynamical mass generation involving strongly-interacting mirror fermions. As a paradigm which has been explicitly shown to predict correctly the weak scale and the weak angle and thus addressing successfully the essence of the hierarchy problem of energy scales (contrary to "supersymmetric" or "extra-dimension" models) and the unification of fundamental forces, it might have already manifested indirectly its validity experimentally, as first noted by the author in 1998, via the bottom-quark forward-backward asymmetry and the related coupling parameter $A\_b$. After a decade during which the particle-physics community was frequently misled by a particularly obscurantist interpretation of the L.E.P./S.L.C. precision data, this approach emerges as a strong candidate for new physics beyond the standard model of elementary particles to be thoroughly tested in the forthcoming high-energy experiments.

**Introduction**

We are all used to taking the physical quantity of mass for granted. Nevertheless, the mystery surrounding its genesis remains hitherto unsolved. Taking a reductive approach, high-energy physicists try to discover the origin of mass of the elementary ingredients of our world. A new high-energy physics experiment based on a proton accelerator named "Large Hadron Collider" (L.H.C.) aims at answering this question which currently dominates elementary-particle physics. Currently under construction, it is scheduled to start operating at the European Organization for Nuclear Research (C.E.R.N.) near Geneva by Spring 2007. If successful, after several months of colliding protons with very high energies against each other, it might create new particles and thus provide data enabling us to unravel some of the most fundamental aspects of mass generation.

To begin with, contemplating on our world consisting of massless ingredients (i.e. particles without mass) might help us realize the importance of solving this question. This is not as absurd as it first sounds, since there are several properties of elementary particles besides mass (by mass here and in the following we mean mass of motionless particles) which render this question meaningful. These properties along with mass dictate the behaviour of particles under the four distinct fundamental interactions detected so far in nature, i.e. the electromagnetic, strong nuclear, weak nuclear and gravitational interactions. They are found to be succinctly described by the mathematical theory of symmetry groups as we will later see, and their study is the main goal of elementary-particle physics.

Consequently, we describe the current view of physicists with regards to mass generation. According to this view, particles acquire their mass by interacting with new heavy particles which we expect to discover in L.H.C. or similar future high-energy experiments. Nevertheless, the particular outcome of these experiments is uncertain, since the set of properties specific to these new particles is the object of an on-going debate. We choose to discuss here only one of the likely outcomes of the L.H.C. experiment which emerged recently via a series of publications by the author as a very attractive possibility and is favored both theoretically and experimentally. This outcome, apart from solving the mystery of mass, will deepen our understanding of nature and reveal a hitherto undetected "mirror world". The concepts of mass and various energy scales, the unification of fundamental forces of interacting particles, and mirror symmetry are thus brought together in an intriguing and instructive theoretical construct.

**Particle masses**

Massless particles? The theory of relativity teaches us that if elementary particles had no mass, they would all travel with the speed of light, i.e. a constant speed about seven orders of magnitude (i.e. seven factors of ten) faster than a car in a highway. Nothing would be able to stay motionless or travel slower or faster than anything else. Then, we have to mention the relevant implications of quantum theory. Quantum theory predicts that, if the electron for instance had no mass, the existence of atoms, i.e. electrons moving around atomic nuclei, with finite size would be impossible.

Why would this happen? Since the size of electron orbits around an atomic nucleus is calculated to be inversely proportional to the electron mass, nuclei would never be able to force massless electrons to turn around them and form atoms, since the radius of their orbits would be theoretically infinite. This would render nearly all branches of natural science, like solid-state physics for instance, obsolete, and the complexity of the world as we know it unconceivable. Under such conditions, the birth of life, and even more so life being conscious of its own existence, would be impossible.

Fortunately, nature chose otherwise. Experiments during most of the 20$^{th}$ century enabled us to discover several matter particles (named electrons, muons, neutrinos, quarks, etc.) widely believed by now to be elementary. Their discovery was almost always followed by a measurement of a non-zero value for their mass. After having analyzed all the fundamental constituents of matter around us, physicists still continue to build experiments which collide particles against each other with increasingly large energies. These collisions transform some of the energy of the initial particles to the masses of particles emerging afterwards, in accordance with relativity theory which relates mass and energy.

Among the collision debris, they try to discover new particles which are much heavier than the presently known ones, even though still far from being visible with a naked eye. They presume that these new particles existed only at the beginning of our Universe, they decayed to the lighter ordinary particles shortly afterwards because of their instability, and they re-emerge for tiny fractions of a second in modern high-energy experiments. Their existence completes the classification of elementary particles in a very symmetric mathematical framework based on their interactions. Some of these particles are the subject of this article, as we will shortly see.

In 1995, experimentalists discovered the heaviest (unstable) particle known so far, named "top quark". They measured its mass to correspond to an energy of around 0.175 TeV. TeV is an energy-measurement unit, and it is nearly twelve orders of magnitude larger than the average energy of a photon inside the light-beams coming from our sun. In addition, it is the energy scale which will be probed in the L.H.C. experiment.

The measurement of the mass of the top quark has left us with a quite complicated mass spectrum of particles – from the lightest neutrino to the heaviest quark - spanning at least eleven orders of magnitude, a puzzle which we still barely understand. Important exception to this are the masses of two particles named "up" and "down" quarks, which are mostly due to the strong nuclear interaction, constitute most of the mass of atoms and molecules and contribute to most of the weight of the objects around us. Elementary-particle physicists introduce in their theories particles which are initially massless. They then add their masses arbitrarily (*via* the Higgs particle described next) in order to fit phenomenological constraints but with no real explanation for their values. The analogy with the situation one century ago when physicists discovered complicated atomic spectra but could not account for them until the advent of quantum theory is quite compelling.

The top-quark mass corresponds to an energy close to the TeV scale, which is the energy not only believed to hold the key to mass generation, but also beyond which the standard model of elementary particle physics is theoretically inconsistent. As a solution, physicists usually introduce *ad hoc* a new heavy (unstable) hypothetical elementary particle not detected yet, which they call "Higgs particle" after Peter Higgs at Edinburgh University who first proposed its existence. Moreover, they assume that the masses of the known particles emerge *via* the "Higgs mechanism" involving interactions with the Higgs particle.

This new particle should have a mass close to the top-quark mass. Unfortunately, it cannot solve satisfactorily the inconsistency of the theory if it is assumed elementary (i.e. non-composite). Even if we initially fix its mass, corrections due to quantum theory give huge contributions to its value and render it too large. These quantum corrections would render our world unacceptably heavier than what is measured experimentally, since the masses of particles are found to be proportional to the mass of the Higgs particle.

To get a rough idea of the size of the problem, if nothing were done to cure the theory, we would predict that the Higgs particle and the top quark for instance weight about the same as a small piece of dust, i.e. about 20 billionths of a kilogram. In other words, they would have a mass corresponding to the "Planck scale", i.e. the energy scale where we expect gravitational interactions to be strong enough to halt the further increase of quantum corrections on mass mentioned above. On the contrary, we find experimentally the top quark weighting seventeen orders of magnitude less! This is a simple way to describe what physicists usually call the "hierarchy problem" of energy scales.

Theorists have found of course mathematical methods enabling them to fix the predicted value of an elementary Higgs mass to acceptable levels by using "extra dimensions" or "supersymmetry". These methods give results consistent with current phenomenology and are exactly solvable. In the absence of relevant experimental data, the above traits lead theorists towards research paths of rather idealistic direction. Nevertheless, it is very important to realize that they stay short of solving the essence of the hierarchy problem. This has not to do just with the question of why the Higgs mass is so small. The deeper question is why it takes this particular value, which also determines the masses of the rest of the elementary particles in our world and which is seventeen orders of magnitude smaller than the Planck scale, and why it is not equal to this scale or of some other arbitrary size.

In the late seventies, Steven Weinberg (now at the University of Texas at Austin) and Leonard Susskind (Stanford Linear Accelerator Center (S.L.A.C.)), adopting a more realistic approach, speculated that the Higgs particle might not be elementary, but a composite one consisting of pairs of new particles [1]. In particular, they proposed the existence of new, so far undetected, strongly-interacting particles with masses close to the TeV scale. Their dynamics, which are usually quite hard to analyze, should be able to produce a composite Higgs mechanism giving mass to ordinary elementary particles. This frees the effective Higgs mass from the large quantum corrections when it is assumed elementary.

Almost two decades of efforts towards such a direction have nonetheless proved frequently problematic [2], since the corresponding theories suffer from several phenomenological disadvantages, especially after the experimental data of the electron colliders L.E.P. and S.L.C. of the early nineties. Moreover, they can not lead to a unified theoretical framework encompassing all elementary particles and interactions between them including gravity, a dream not foreign to Parmenides monism and Kant's *Critic of pure logic* (1787): "Logic presupposes the unity of various forces, because particular natural laws stand under more general ones, and the sparing of principles is not only an economical postulate of logic, but also inner law of nature". This situation changed dramatically however during the last three years by invoking the "mirror world" described in the following.

**The mirror world**

Particle-physics experiments in the mid-fifties indicated that nature did not respect parity symmetry, i.e. it was not left-right symmetric. This simply means that if particles are reflected in a mirror, the reflected image has an orientation or displays properties which either cannot be found or are very rare in particles of the real world. We know already numerous examples of left-right asymmetries in other natural sciences. As in chemistry, where chiral molecules like L-Dopa, used as a cure for Parkinson's disease, have different properties than their mirror partners and formed the basis of the 2001 Nobel prize in chemistry. Or like DNA molecules, which form almost always left-handed helices. And going further to life sciences and to the human bodies, where the position of the various organs clearly violates left-right symmetry.

To understand the asymmetry now as it appears in the microcosmos in simple terms, imagine two elementary matter particles with the same mass but opposite directions of motion and opposite electric charges, one coming from the left and one from the right, colliding against each other. On average, the debris of the collision flying to the left is measured to be different from the one flying to the right. This asymmetry is commonly referred to as "forward-backward asymmetry".

T.D. Lee (Columbia) and C.N. Yang (now at SUNY at Stony Brook) explained the phenomenon of parity-symmetry violation in a landmark paper published in 1956 [3]. A new ingredient is needed in order to understand their proposal, and this is called "spin". Spin is a quantity of quantum-theoretical nature sharing only a few properties with classical spin. Matter particles can be either right- or left-handed, according to whether their spin has the same or opposite direction to their motion. Watching these in a mirror, we would have the opposite picture consisting of "mirror particles", since it would transform left-handed and right-handed particles to each other (Diagram 1). This is due to the fact that spin and direction of motion have opposite properties under mirror reflection.

Lee and Yang assumed correctly that only left-handed and not right-handed particles "feel" the weak nuclear interaction, i.e. one of the four known forces. (We have to exchange "left" and "right" above when speaking of anti-particles, i.e. particles with the same properties but opposite electric charge than ordinary particles.). Conversely, mirror right-handed particles (and mirror left-handed anti-particles) feel the weak nuclear interaction, whereas mirror left-handed particles (and mirror right-handed anti-particles) do not. Mirror particles are predicted to exhibit a forward-backward asymmetry of opposite sign than ordinary particles (Diagram 2). This is due to the fact that weak nuclear interactions are not invariant under parity (mirror reflection), but invariant only under the combined transformations of charge conjugation (interchange of particles with their anti-particles), parity and time reversal.

Experiments so far have given no indication for the real existence of mirror particles. How can nature in its most fundamental manifestation be so asymmetric? Lee and Yang first speculated in that 1956 paper on the existence of a "mirror world", i.e. on new particles being the mirror partners of ordinary particles, in order to restore parity symmetry in a broader sense. This is reminiscent of the (experimentally proven since several decades now) existence of anti-particles predicted by P.A.M. Dirac, even though it is still unclear why our world consists so asymmetrically almost exclusively of particles and not of anti-particles.

The fourfold complex consisting of particles/anti-particles/mirror particles/mirror anti-particles would complete in a robust and symmetric framework our understanding of elementary particles. Such fourfold classification schemes perfecting the classical dialectic duality have frequently appeared in human thought, from Pythagoras' tetraktys and the four elements of the ancients to Kant's ontology, and from physics to psychology; they intrigued W. Pauli and lead C.G. Jung to write " Quaternity is an archetype of almost universal occurrence".

What would happen now if we tried to combine the two ideas exposed above, i.e. the existence of mirror particles with a composite Higgs particle? What if the hypothetical Higgs particle is not elementary but consists of a pair of mirror particles?

The results of this combination have proven to be quite intriguing. In a series of papers published recently, we showed that the existence of (unstable) mirror particles, not discovered yet, with masses around the TeV scale can produce successfully an effective composite Higgs mechanism [4]. (A previous very interesting attempt along similar lines predicted particles about seven orders of magnitude too heavy [5]).

These particles give a solution to several important problems currently plaguing high-energy physics and are consistent with the L.E.P./S.L.C. data. They were named "katoptrons" by the author in 2000 [4] after the Greek word for mirrors, since it is important to differentiate them from the mirror particles of Lee and Yang: apart from "feeling" the four known interactions, we assume that katoptrons interact with each other also *via* a new interaction which is very strong at the TeV energy scale. This is reminiscent of the strong nuclear force giving mass and binding quarks together to form protons and neutrons, only it is operative at energies which are three orders of magnitude larger. This gives katoptrons large dynamical masses, creates composite Higgs particles, and explains why katoptrons are unstable and so much heavier than ordinary matter particles.

To be more precise, under the three symmetry groups known as SU(3), SU(2) and U(1) describing the strong nuclear, weak nuclear and electromagnetic forces respectively, particles are characterized by ("carry") three quantum "numbers" that dictate their corresponding interactions: On one hand, ordinary left-handed quarks (i.e. up, down, top, bottom *etc*.) and leptons (i.e. electrons, muons, neutrinos *etc*.) carry the numbers (**3**, **2**, 1/3) and (1, **2**, -1) respectively, while left-handed antiquarks and antileptons carry the numbers (**3\***,1, -4/3 and 2/3) and (1, 1, 0 and 2) respectively. In the above, **3** and **2** indicate among other things the number of different particle "species" with respect to the corresponding interaction, 1 under SU(3) or SU(2) means that the particle in question does not have this interaction, and the U(1) quantum numbers are directly related to the usual electromagnetic charges. Three copies ("generations") of these particles have been found so far, with identical quantum numbers but different masses. Never before in human history has such a small set numbers contained so much information which is experimentally confirmed, and which forms the basis of the theory known as "standard model".

On the other hand, we predict that the above particles have right-handed mirror partners with the same quantum numbers, and which in addition carry a new quantum number **3** dictating their interactions (and the number of "species") corresponding to a new force described by a symmetry group SU(3)′. The fact that ordinary particles come in three generations whereas their mirror partners are grouped under the number **3** of SU(3)′ indicates that an initial (discrete) symmetry between these two kinds of particles is now broken. This might have occurred if our world had at the initial moments of its creation nine instead of the present three space dimensions [4].

We have now at last a definite theoretical framework in order to explore and explain in principle more rigorously the wide spectrum of the masses of elementary particles, which in this context is generated by the aforementioned interactions.

Moreover, we can solve the essence of the hierarchy problem, since we do not assume *a priori* the TeV energy scale as the models with elementary Higgs particles do. The new strong interaction we introduce allows us to derive the TeV scale instead by simply assuming that ***it can be unified with the other interactions*** at a large energy scale close to the Planck scale called "unification scale". This scale is the energy where the strengths of all interactions are equal to each other.

How is this done? Just as quantum effects influence particle masses, they also influence the strength of interactions at different energies. The strengths of the strong nuclear interaction and the new strong interaction described by SU(3)′ decrease with increasing energy, and they both reach a common point with the other interactions at the unification scale.

In particular, with the unification assumption the ratio of the unification scale to the TeV scale is correctly predicted to be approximately equal to the $\exp(1/a) \sim$ thirteen orders of magnitude, where $a \sim 1/30$ measures the strength of the weak nuclear interaction [4]. The unification of forces guarantees in addition that the weak angle, a quantity measuring the relative strength of the weak nuclear and electromagnetic interactions, is predicted in accordance with experiment. Moreover, at the unification energy scale all symmetries describing the various interactions are embedded in one larger symmetry. Left- and right-handed particle species are then grouped under the same quantum "number" corresponding to this symmetry, adding to the simplicity of the theory.

Therefore, the existence of new strongly-interacting heavy particles at the weak scale, named katoptrons, and unification of all interactions in nature within a common theoretical framework are concepts for the first time not mutually exclusive in a theory which is phenomenologically viable. Kant's thoughts cited before are thus brought back to relevance within a particular framework involving the "mirror world".

**Future prospects**

Obviously, beyond checking a physical theory for its theoretical consistency and mathematical beauty, we must also subject it to the difficult tests of experimental scrutiny. We might have already measured indirect signatures of katoptrons *via* effects described by quantum theory which mix them with ordinary particles in proportion to the values of their masses. For instance, the experimentally measured value of the coupling parameter $A_b$ of the second heaviest elementary particle known so far, the "bottom quark", deviates from the value predicted by the usual models of elementary Higgs particles with a statistical confidence of 99.7%, a 3 standard deviations (s.d.) effect!

In particular, the L.E.P. results give $A_b = 0.893 \pm 0.022$ and the combined L.E.P./S.L.C. results give $A_b = 0.901 \pm 0.013$, which deviate 3.2 and 2.6 s.d. respectively from the theoretically predicted value of 0.935, and which are mutually consistent with $R_b$, another important quantity related to the bottom quark [6]. Contrary to these models, katoptron theory is unique in explaining this deviation [4]. In addition, recent results from the L.E.P. experiment at C.E.R.N. indicate that the Higgs particle is heavier than previously thought, which can be viewed as another hint that it is not elementary.

Apart from such indirect effects which can hardly be conclusive, full experimental verification of katoptron theory may come only after the direct production of the new heavier particles it predicts. On Summer 2007 at the earliest, we might have enough data from the L.H.C. experiment near Geneva to conclude on their existence. This will settle *a fortiori* the heated debate on which types of new particles have masses lying near the TeV energy scale and are responsible for the masses of ordinary particles [7].

What will we have achieved after this effort? We believe that katoptrons were first created with the beginning of our world, only to live less than a billionth of a second and then decay. They still leave some traces behind and affect our world indirectly by giving mass to ordinary elementary particles *via* a composite Higgs mechanism. We expect that the next high-energy experiments, unless the "Higgs mechanism" hides something different, will "resuscitate" this mirror world to become more than a mere mirage after billions of years of only virtual existence.

Once again however, katoptrons will not be able to reach their eternal entelechy, since they are inherently unstable. They will reemerge in our world for much less than a billionth of a second, and they will subsequently decay to ordinary particles shortly after their creation. High-energy physicists, seeming more like space-time "archaeologists" nowadays, are eagerly awaiting these brief "awakenings" in order to draw sufficient experimental data enabling us to grasp a firmer handle on the genesis of mass.

# Diagrams

*Diagram 1*: a. Description of left- or right-handedness, the property of elementary particles (depicted as small circles) which helps the explanation of parity symmetry violation. b. Below, the same picture as seen from a mirror. The mirror shows a situation not seen experimentally with real particles yet.

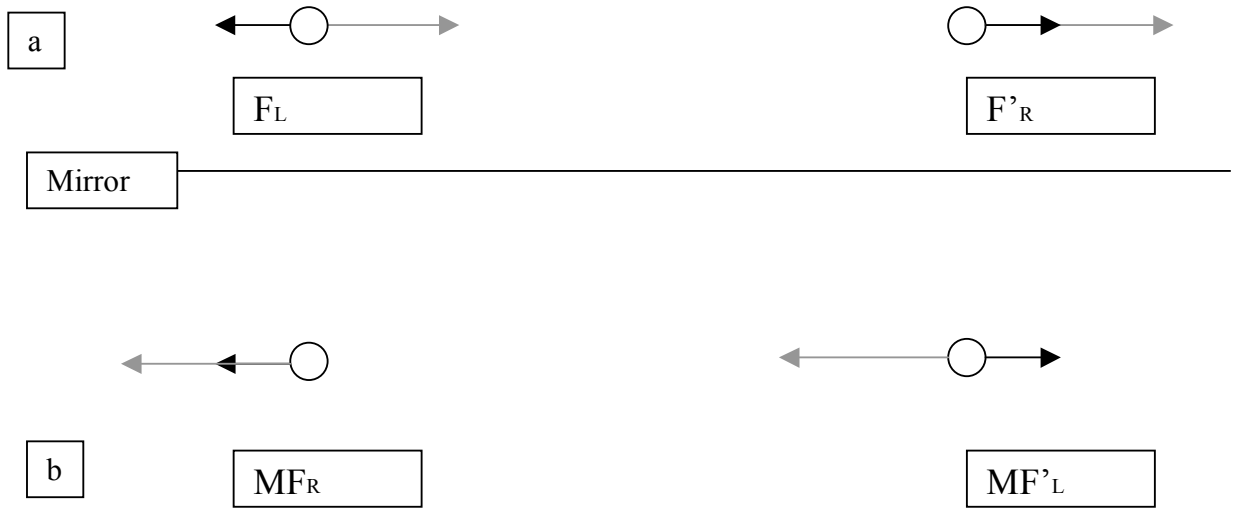

Black arrow: spin, invariant under mirror reflection
Gray arrow: direction of motion, reversed under mirror reflection

$F_L$: left-handed particle (feels weak nuclear interaction)

$F'_R$: right-handed particle (does not feel weak nuclear interaction)

$MF_R$: right-handed particle (feels weak nuclear interaction, not discovered yet), mirror partner of $F_L$

$MF'_L$: left-handed particle (does not feel weak nuclear interaction, not discovered yet), mirror partner of $F'_R$

The anti-particles of the above particles have the opposite left-right properties under the weak nuclear interaction, i.e. a left-handed anti-particle and a right-handed mirror anti-particle for instance do not feel the weak nuclear interaction and *vice-versa*.

*Diagram 2*. A schematic exemplification of the violation of parity symmetry. We consider a very symmetric initial situation with matter particles all travelling with the same energy close to the TeV scale. a. A particle colliding against its anti-particle prefers on average to continue towards its initial direction after the collision. This leads to a "forward-backward" asymmetry. b. On the contrary, mirror particles tend on average to reverse their direction after interacting with their anti-particles, leading to a "forward-backward" asymmetry of opposite sign.

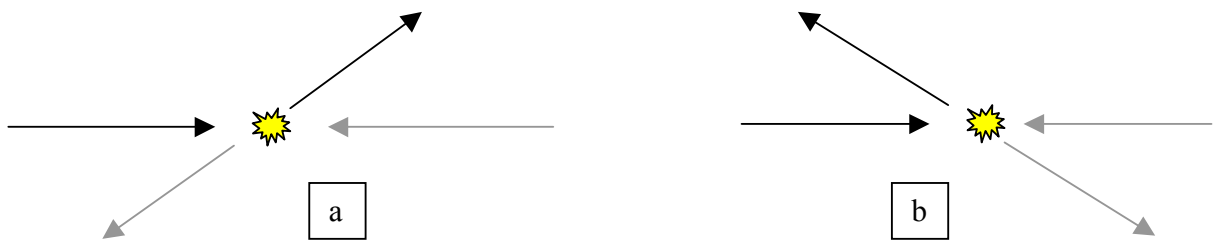

Black arrow: direction of particle
Gray arrow: direction of anti-particle

Black arrow: predicted direction of mirror particle (not discovered yet)
Gray arrow: predicted direction of mirror anti-particle (not discovered yet)